\documentclass[sigplan,screen]{acmart}
\usepackage{algorithm}
\usepackage{algorithmic}
\usepackage{listings}
\usepackage{xcolor}  
\usepackage{tabularx}
\usepackage{array}
\usepackage{soul}
\newcolumntype{Y}{>{\raggedright\arraybackslash}X}
\usepackage{enumitem}
\usepackage{makecell}
\usepackage{todonotes}
\usepackage{graphicx}
\usepackage{subcaption}
\usepackage{amsmath}
\usepackage{xspace}
\floatstyle{plain}
\restylefloat{listing}
\usepackage{xcolor}
\usepackage[debug=true,frozencache=true,cachedir=_minted]{minted}
\DefineVerbatimEnvironment{MintedVerbatim}{Verbatim}{}
\usepackage{tcolorbox}
\tcbuselibrary{skins,listings,minted,breakable}

\definecolor{codebg}{HTML}{F7F7F8}
\definecolor{codeframe}{HTML}{D0D0D0}
\definecolor{codetitle}{HTML}{ECECEC}
\definecolor{codetext}{HTML}{202020}

\definecolor{linenumbergray}{HTML}{A0A0A0}

\newtcblisting{vscode}[2][]{%
  listing engine=minted,
  minted language=#2,
  minted options={
    style=vs,
    fontsize=\scriptsize,
    autogobble,
    breaklines,
    breakanywhere=false,
    breaksymbolleft={},
    breaksymbolright={},
    tabsize=2,
    encoding=utf8,
    linenos,
    numbersep=6pt
  },
  listing only,
  enhanced,
  colback=codebg,
  frame hidden,
  boxrule=0pt,
  coltext=codetext,
  arc=0.8mm,
  outer arc=0.8mm,
  left=1.5mm,
  right=1.5mm,
  top=1.2mm,
  bottom=1.2mm,
  boxsep=0pt,
  before skip=4pt,
  after skip=4pt,
  title={#1},
  colbacktitle=codetitle,
  coltitle=black,
  fonttitle=\ttfamily\scriptsize,
  boxed title style={
    boxrule=0pt,
    arc=0.8mm,
    outer arc=0.8mm
  },
  attach boxed title to top left={
    xshift=0mm,
    yshift=-\tcboxedtitleheight
   }
}
\AtBeginDocument{%
  }

\copyrightyear{2027}
\acmYear{2027}
\setcopyright{cc}
\setcctype{by}
\acmConference[ASPLOS '27]{Proceedings of the 32nd ACM International Conference on Architectural Support for Programming Languages and Operating Systems, Volume 1}{April 11--15, 2027}{Heraklion, Greece}
\acmBooktitle{Proceedings of the 32nd ACM International Conference on Architectural Support for Programming Languages and Operating Systems, Volume 1 (ASPLOS '27), April 11--15, 2027, Heraklion, Greece}
\acmDOI{10.1145/3845814.3850077}
\acmISBN{979-8-4007-3012-2/2027/04}
\ccsdesc[500]{Computer systems organization~Architectures}
\ccsdesc[300]{Hardware~Memory and dense storage}

\newcommand{\matr}[1]{\ensuremath{\mathbf{#1}}} 

\theoremstyle{definition}
\newtheorem{definition}{Definition}[section]

\begin{document}

\title[]{Programmable Data Layouts via On-The-Fly Data Transformation}
\author{Cole Strickler}
\email{colestrickler@ku.edu}
\affiliation{%
  \institution{University of Kansas}
  \city{Lawrence}
  \state{Kansas}
  \country{USA}
}

\author{Renato Mancuso}
\email{rmancuso@bu.edu}
\affiliation{%
  \institution{Boston University}
  \city{Boston}
  \state{Massachusetts}
  \country{USA}
}

\author{Manos Athanassoulis}
\email{mathan@bu.edu}
\affiliation{%
  \institution{Boston University}
  \city{Boston}
  \state{Massachusetts}
  \country{USA}
}

\author{Heechul Yun}
\email{heechul.yun@ku.edu}
\affiliation{%
  \institution{University of Kansas}
  \city{Lawrence}
  \state{Kansas}
  \country{USA}
}

\renewcommand{\shortauthors}{Strickler et al.}

\begin{abstract}


The performance of modern computing systems is heavily bottlenecked by the widening gap between processing throughput and main memory latency, a problem exacerbated when an application's logical data access patterns misalign with the physical layout of data in memory. To address this, we propose a hardware-software co-design framework comprising the Data Transformation Language (DTL) and the Data Transformation Unit (DTU). DTL allows programmers to concisely express complex data layout reorganizations, which the compiler translates into efficient hardware routing configurations. The DTU, a hardware engine logically situated between the Last-Level Cache (LLC) and main memory, transparently evaluates these configurations on-the-fly to gather scattered data elements into dense, logically contiguous cache lines. By converting explicit layout transformations into virtual views, this approach eliminates the need for explicit data materialization in software. We evaluate the DTU within a quad-core RISC-V SoC using FireSim, demonstrating substantial improvements in execution speed and memory efficiency. 


    
\end{abstract}

\keywords{data transformation, data layout, cache locality, memory systems, hardware-software co-design}

\maketitle

\section{Introduction}

Data movement between processing elements (CPUs, GPUs, and other accelerators) and main memory is a major bottleneck for both performance and energy efficiency in modern systems. This challenge has become more acute with the proliferation of data-intensive applications across machine learning, databases, and high-performance computing. Compounding the problem, memory technology scaling has lagged behind the throughput growth of contemporary processing units~\cite{memory-wall-reflections, wulf1995hitting, aimemorywall}, creating a pronounced memory performance gap. To mitigate long-latency accesses, modern systems rely on multi-level caches, complex prefetching strategies~\cite{event-trigger-prefetch, prefetch-dependence-graph, prefetch-stream-based, prefetching-synthesis}, and increased memory-level parallelism (MLP). However, the effectiveness of these mechanisms highly depends on how data is laid out in memory.

In many applications, the logical organization of data does not align with the layout required for efficient utilization of the memory hierarchy. As a result, software often performs explicit data layout transformations, materializing multiple versions of the same data object optimized for different access patterns~\cite{Arulraj2016}. These transformations incur substantial overheads in data movement, memory footprint, and execution time. Furthermore, they entangle algorithmic logic with layout and hardware-specific optimizations, complicating programmability and maintainability.

Existing approaches attempt to address this challenge through both software and hardware techniques. Compiler-driven frameworks and tensor systems~\cite{tvm,xla,halide} typically generate a single materialized layout optimized for a target computation. Supporting multiple layouts requires duplicating data, performing additional transformations, or introducing complex index mappings~\cite{graphene-IR}, all of which incur additional memory and runtime overhead. The Processing-Near-Memory (PNM) and Processing-In-Memory (PIM) approaches~\cite{mutlu2025PIM, pnm_past_present_future} reduce data movement by bringing computation closer to data. Unfortunately, these approaches often restrict the class of supported computations and require significant architectural and programming model modifications.

To address these challenges, recent works have tried to build \emph{on-the-fly} transformation engines, which focus on moving only useful data into the memory hierarchy and providing software with ideal locality. These methods have targeted application specific access patterns including databases~\cite{roozkhosh2023edbt-rme,strickler2025adms-rme,rel-fabric} and tensor operations~\cite{tme-paper}. Our design builds upon these approaches, but looks to overcome domain specificity to make cache line contents generally \emph{programmable}, enabling a unified abstraction applicable across diverse workloads.

We propose a hardware–software co-design framework comprising the Data Transformation Language (DTL) and the Data Transformation Unit (DTU). DTL provides a concise, C-like specification that allows programmers to express how logical data coordinates map to physical memory addresses. The DTU, a hardware engine written in Chisel HDL~\cite{chisel2012} and situated between the LLC and main memory, evaluates these compiler-generated mappings on-the-fly. By dynamically gathering scattered positional data into dense, logically contiguous cache lines, the DTU achieves ideal cache locality. Crucially, this framework converts explicit layout transformations into virtual views, eliminating materialization overheads, reducing memory footprint, and operating entirely without modifications to existing computation kernels.

In summary, we make the following contributions:

\begin{itemize}[leftmargin=12pt]

\item We introduce a formal abstraction that decouples logical data access from physical memory layout, enabling layout transformations to be expressed as programmable address mappings.
\item We design and implement the \textit{Data Transformation Unit} (DTU), a hardware engine that constructs programmable cache lines on-the-fly.
\item We develop the \textit{Data Transformation Language} (DTL) and compiler infrastructure to express and realize transformation mappings.
\item We provide a full-system implementation---including synthesizable hardware, compiler toolchain, OS driver and runtime library---and evaluation on a RISC-V SoC in FireSim, demonstrating that programmable data transformation is both practical and effective across a range of workloads.

\end{itemize}

\section{Background}\label{sec:background}

In this section, we provide background on the importance of data layouts and on recent work on domain-specific on-the-fly data transformation engines.

\subsection{The Data Layout Challenge}\label{sec:bg_datachallenge}
Modern computing systems rely heavily on multi-level cache hierarchies and hardware prefetchers to bridge the widening performance gap between the CPU and main memory ~\cite{wulf1995hitting}. Caches exploit spatial and temporal locality by fetching data from main memory in fixed-size contiguous blocks, known as cache lines. Hardware prefetchers complement this by monitoring memory access streams and proactively fetching adjacent or predictably strided cache lines before the CPU explicitly requests them~\cite{jouppi1990improving}.

However, these mechanisms are highly optimized for dense, predictable memory access patterns. While data-intensive applications have been heavily optimized to generate such access patterns (e.g., column-stores for data analytics \cite{abadi2013design, Boncz2002}), providing the optimal layout for ad hoc queries remains an open question \cite{Arulraj2016, roozkhosh2023edbt-rme}, because the access patterns remain unpredictable.
In a similar vein, data-intensive applications involving high-dimensional tensor manipulations often exhibit data traversals that are misaligned with the physical memory layout \cite{Springer2017HPTT}. This misalignment destroys spatial locality. The CPU is forced to fetch full cache lines only to consume a small fraction of the loaded elements, leading to poor cache line utilization and inflated working set sizes. Furthermore, the resulting irregular, large-stride access patterns easily defeat traditional hardware prefetchers, leading to frequent cache misses, saturated memory bandwidth, and stalled compute pipelines.

The need for well-aligned, densely packed data is even more pronounced when using Single Instruction Multiple Data (SIMD) instructions and other hardware accelerators that exploit data parallelism~\cite{Eichenberger2004VectorizationSIMDAA}. These vector execution units require contiguous memory operands to achieve peak computational throughput. The performance degradation caused by irregular data is so severe that highly optimized software libraries frequently resort to explicitly duplicating data into a reorganized, contiguous layout prior to execution \cite{chellapilla2006high, abadi2013design}. Despite the execution time required for this data movement and the memory footprint overhead incurred by materializing intermediate views (e.g., \textit{im2col} for convolutions), this software-driven data duplication remains a necessary trade-off to unlock optimal compute performance.
\vspace{-4pt}
\subsection{Domain-Specific Data Transformation Engines}\label{sec:bg_domain}

To address the severe performance penalties associated with data layout mismatches, recent research has explored hardware-accelerated, on-the-fly data reorganization. These approaches introduce specialized transformation engines situated between the CPU and main memory. By intercepting memory requests, these engines dynamically gather non-contiguous data elements from the backing store and repackage them into dense, optimal cache lines before delivering them to the processor's cache hierarchy.

Notable examples of this paradigm include Relational Memory (RME) \cite{roozkhosh2023edbt-rme} and the Tensor Memory Engine (TME) \cite{tme-paper}. RME is specifically tailored for relational database systems; it dynamically extracts and packs target columns from a base row-store layout, presenting contiguous columns to the CPU~\cite{roozkhosh2023edbt-rme, strickler2025adms-rme}. Similarly, TME is designed to accelerate tensor operations by executing on-the-fly multi-dimensional slicing, transposing, and folding, feeding dense data streams to the processing pipelines.

While both RME and TME successfully demonstrate the benefits of decoupling access pattern from physical layout, they are fundamentally constrained by their domain-specific designs. RME relies on hardware logic tailored specifically to database geometry (e.g., row sizes and column offsets), whereas TME is tightly coupled to fixed tensor stride manipulations. This lack of generality means that neither engine can be easily adapted to accelerate arbitrary layout transformations across broader application domains. Consequently, there remains a critical need for a fully programmable, generalized data transformation unit capable of supporting arbitrary, on-the-fly memory re-mappings.

\section{Motivating Example}
\label{sec:motivating-example}
To motivate on-the-fly data transformation, consider a naive matrix multiplication $\matr{C}=\matr{A}\matr{B}$, where $\matr{A}$ and $\matr{B}$ are $16\times16$ matrices stored in row-major order. Assume each matrix element occupies 4 bytes and the cache-line size is 64 bytes. A conventional matrix-multiplication kernel computes each element of $\matr{C}$ by taking the dot product of a row of $\matr{A}$ and a column of $\matr{B}$.

The accesses to $\matr{A}$ have good spatial locality because consecutive elements of a row are contiguous in memory. The accesses to $\matr{B}$, however, are strided. Since each row of $\matr{B}$ contains 16 elements, it occupies exactly one 64-byte cache line. Consequently, two consecutive elements of the same column of $\matr{B}$ are separated by 64 bytes and reside in different cache lines. A traversal of one column therefore touches 16 different cache lines while consuming only one 4-byte element from each line. This mismatch between the stored layout of $\matr{B}$ and the access pattern of the computation results in poor spatial locality and can generate substantially more cache and memory traffic than the amount of useful data consumed.

A common optimization is therefore to transpose $\matr{B}$ before performing the multiplication. After transposition, the elements that formed a column of $\matr{B}$ form a contiguous row of $\matr{B}^{\intercal}$. The multiplication can then traverse both $\matr{A}$ and $\matr{B}^{\intercal}$ sequentially, making effective use of each fetched cache line. In this example, the 16 elements required from $\matr{B}$ for a dot product occupy 16 separate cache lines in the original layout, but exactly one 64-byte cache line in the transposed layout.

However, this software-based transposition requires additional computation to rearrange the matrix, additional data movement to read and rewrite its elements, and additional memory capacity to store the transformed intermediate representation---all of which are undesirable overheads.  

Our approach avoids these costs by performing the transformation in hardware on the fly, without materializing the transposed matrix in physical memory. We return to this example throughout the paper to illustrate the design and operation of the system.
The remainder of the paper develops the abstraction, programming interface, and hardware mechanisms that enable this capability. 
\section{Data Transformation} \label{sec:data-transformation}

This section introduces \emph{data transformation} and its underlying theoretical framework. The fundamental principle is the decoupling of the application's \emph{logical view} (how memory objects are addressed) from its \emph{positional view} (how objects are serialized in memory), defined formally as follows:


\begin{definition}[Positional Address Space]
A \emph{positional address space} is an address space in which data addresses directly encode the position in memory where the data item to be accessed is stored. 
Once an object is materialized in memory, it has a unique positional address space. Data addresses in this space are henceforth called \emph{positional data addresses} or simply \emph{positional addresses}.
\end{definition}

\begin{definition}[Logical Address Space]
A \emph{logical address space} is an address space where the layout  progression of addresses reflects the order of data items as required by the semantics of the algorithm or application. Data addresses in this space are henceforth called \emph{logical data addresses} or simply \emph{logical addresses}.
\end{definition}


For example, considering the matrix multiplication example in Section 3, the original matrix $\matr{B}$ resides in a positional address space following the way it is stored in memory, whereas the proposed approach enables the definition of a new logical address space to directly access $\matr{B}^{\intercal}$.

Note that the concepts of logical and positional addresses are orthogonal to the concepts of \emph{virtual} and \emph{physical} addresses. Indeed, from our perspective, a virtual-to-physical translation only establishes a 1:1 mapping of positional data addresses. As such, both positional and logical address spaces can be defined in virtual or physical address space.

In traditional computing systems using positional addresses, the burden of computing the correct position of each data item falls on the application logic. Because the architectural design of memory hierarchy components (e.g., caches, interconnects, and memory controllers) favors cases where the access pattern matches the positional view, a mismatch often results in a significant performance impact. Addressing the inefficiencies arising from the mismatch has been the focus of extensive literature, as discussed in Section~\ref{sec:background}.

In this paper, we propose a programmable,
\textbf{hardware-assisted data transformation} approach to address
mismatches between an application's data access pattern and the
positional layout of its data objects. 

\begin{definition}[Data Transformation]
\emph{Data transformation} is the on-the-fly translation of logical data addresses accessed by an application into their corresponding positional data addresses. This allows data items that are positionally scattered to appear as logically contiguous to the application, without requiring explicit data re-layout in positional space. 
\end{definition}


Leveraging data transformation enables the proactive assembly of multiple useful data items from non-contiguous locations into a single, contiguous $l$-byte cache line. Given an item size of $d$ bytes, constructing one such cache line requires gathering $l/d$ items.




\subsection{Logical to Positional Mapping}


Consider a $n$-dimensional data object (tensor) $\chi^{D_n\times D_{n-1}\times \dots\times D_1}$ following the layout needed by the application semantics, where $\{D_n,D_{n-1},\dots,D_1\}\in\mathbb{Z}^+$. Any logical address for a data item in $\chi$ has a unique set of \emph{logical coordinates}. Abusing a C-like notation, it is easy to see that the logical address corresponding to the element $\chi[x_n][x_{n-1}]\ldots[x_1]$ has logical coordinates $(x_{n},x_{n-1},\dots,x_1)$. Furthermore, we note that the mapping between logical coordinates and logical addresses is always reversible. 

First, the \emph{index function} $L$ is used to compute the logical address corresponding to logical coordinates $(x_n,x_{n-1},\dots,x_1)$:\\
\begin{equation}\label{eq:linear-index}
\begin{aligned}
    L(x_n,x_{n-1},\dots,x_1) = \sum_{i=1}^{n}x_iS_{D_i},
\end{aligned}
\end{equation}
where $S_{D_i}$ are dimension strides defined as follows:
\[
S_{D_1} = 1,\quad 
S_{D_i} = D_{i-1} S_{D_{i-1}} \quad \text{for } i > 1
\]


Next, given a logical address, the \emph{inverse index function} $L^{-1}$ recovers the full set of logical coordinates from a logical linear address:

\begin{equation}\label{eq:rev-linear-index}
L^{-1}(\mathrm{addr}) =
\begin{cases}
\begin{aligned}
x_n &= \mathrm{addr} \mathbin{//} S_{D_n} \\
x_{n-1} &= (\mathrm{addr} \bmod S_{D_n}) \mathbin{//} S_{D_{n-1}} \\
x_{n-2} &= ((\mathrm{addr} \bmod S_{D_n}) \bmod S_{D_{n-1}}) \mathbin{//} S_{D_{n-2}} \\
&\vdots \\
x_1 &= (\mathrm{addr} \bmod S_{D_{n}}) \dots \bmod S_{D_2} \mathbin{//}S_{D_{1}}
\end{aligned}
\end{cases}
\end{equation}


Once the logical coordinates have been recovered, the corresponding positional address of the data item can be computed using a function $F(x_n,x_{n-1},\dots,x_1)$ with the following interface:
\begin{equation}\label{eq:F}
    F: ({\mathbb{Z}^{+}})^n \to \mathbb{Z}^+.
\end{equation}

\begin{figure}[t]  
  \includegraphics[width=0.48\textwidth]{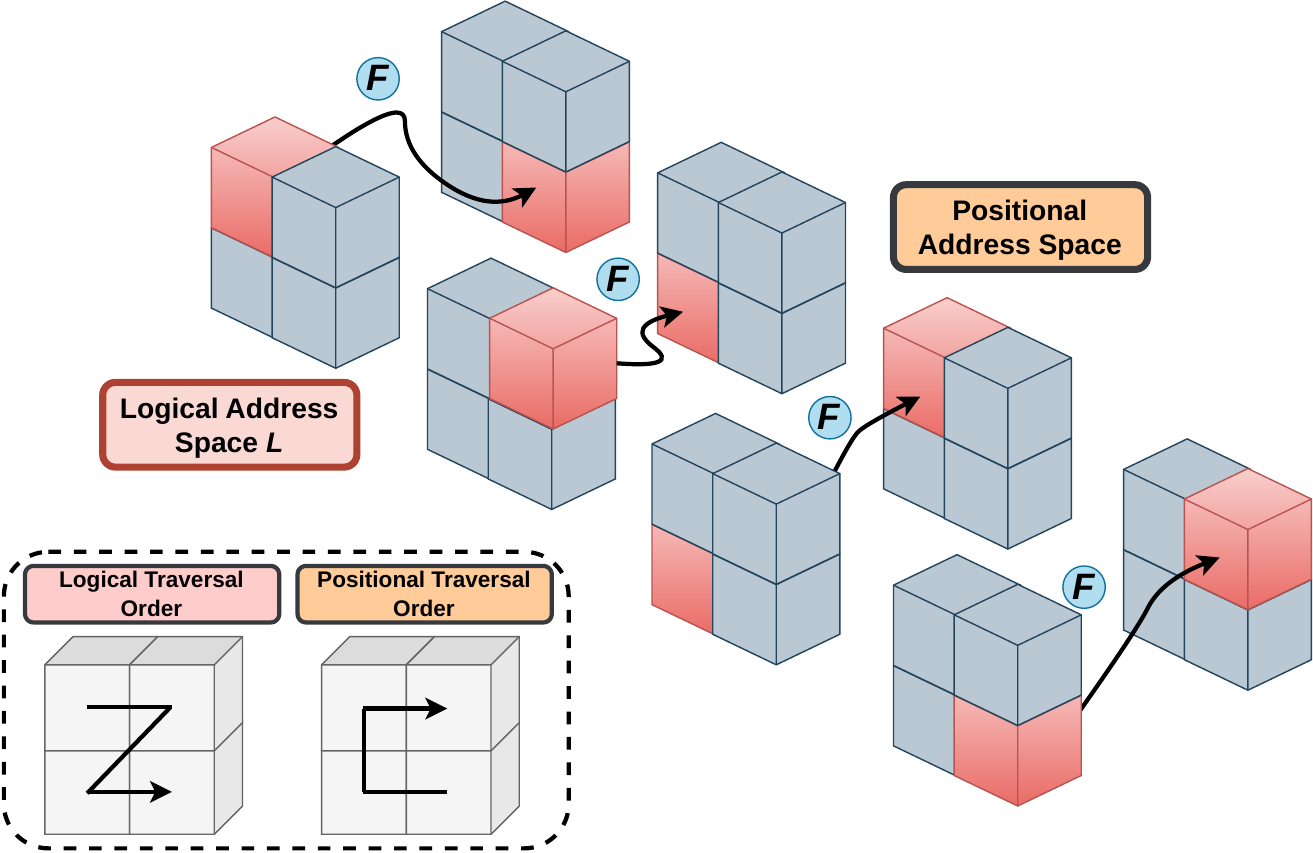}
  \caption{$F$ maps a logical coordinate to a positional address.}
  \label{fig:semantic-decoupling}
\end{figure}

Figure~\ref{fig:semantic-decoupling} illustrates this. In theory, $F$ can be any function of the form given in Eq.~\eqref{eq:F}. In practice, however, the definition of $F$ is constrained by hardware considerations such as available memory, pipeline latency, functional unit composition, and available pipeline depth/width. These constraints determine the class of functions that can be efficiently realized in hardware. The pipeline structure can be instantiated and parameterized at design time to include desired function subclasses. $F$ is not required to preserve contiguity, stride regularity, dimension ordering, or any other property of the original layout. This decoupling of logical coordinates from positional placement is what enables complex re-mappings and, in particular, the construction of dense, high-utility cache lines.

\noindent\textbf{Example.} Consider the matrix $\matr{B}$ introduced in Section~\ref{sec:motivating-example}, which is allocated with dimensions $16 \times 16$ and stored positionally in row-major order. Suppose we request a logical view of $\matr{B}^\intercal$ over this backing store. When accessing the $55^{\text{th}}$ item of $\matr{B}^\intercal$ in logical space, its coordinates $(x_2, x_1)$ are recovered by applying $L^{-1}(55) = (55 // 16, 55 \bmod 16) = (3, 7)$, corresponding to element $\matr{B}^\intercal[3][7]$. The mapping function $F(x_2, x_1) = 16x_1 + x_2 = 115$ then computes the corresponding positional address of $\matr{B}[7][3]$ in the original matrix $\matr{B}$.



\subsection{Efficient Cache Line Packing}

Recall that data transformation allows filling each cache line of size $l$ with $l/d$ items that are sequential in logical address space. Sequential traversal in logical space can be performed efficiently without repeatedly applying $L$. Indeed, once the set of logical coordinates $(x_n, x_{n-1},\ldots, x_1)$ is recovered, the coordinates of the subsequent data item can be computed using the \emph{coordinate successor function} defined in Alg.~\ref{alg:next-coord}.


\begin{algorithm}[htp]
\caption{Coordinate Successor Function $N$}
\label{alg:next-coord}
\small   
\begin{algorithmic}[1]
\STATE \textbf{Input:} $\vec{C_j} = \{x_n,x_{n-1},\dots,x_1\}$ with $0 \le x_i < D_i$
\STATE \textbf{Output:} $\vec{C}_{j+1} =\{x_n,x_{n-1},\dots,x_1\}$
\FOR{$i = 1$ \TO $n$}
    \STATE $x_i \leftarrow x_i + 1$
    \IF{$x_i < D_i$}
        \STATE \textbf{break}
    \ELSE
        \STATE $x_i \leftarrow 0$
    \ENDIF
\ENDFOR
\STATE \textbf{return} $(x_n,x_{n-1} \dots, x_1)$
\end{algorithmic}
\end{algorithm}

\begin{algorithm}[htp]
\caption{Data Transformation Algorithm}
\label{alg:data-transformation}
\small  
\begin{algorithmic}[1]

\STATE \textbf{Input:} $logical\_addr =$ logical linear address
\STATE \textbf{Input:} $l =$ cache line size
\STATE \textbf{Input:} $d=$ data size
\STATE \textbf{Output:} $\vec{P}=$ packed cache line
\STATE $\vec{P} = \langle \rangle$
\STATE $\vec{coords} = L^{-1}(logical\_addr)$
\STATE $\text{needed\_items} = l / d$ 
\STATE $i = 0$
\WHILE{$i < {\mathrm{needed\_items}}$}
    \STATE positional\_address $= F(\vec{coords})$
    \STATE data = \texttt{read\_memory}(positional\_address, $d$)
    \STATE $\vec{P} \,\texttt{.push\_back}(data)$
    \STATE $\vec{coords}=N(\vec{coords})$
    \STATE $i = i + 1$
\ENDWHILE
\STATE \textbf {return} $\vec{P}$
\end{algorithmic}
\end{algorithm}

Using $L^{-1}$, $F$, and $N$, we can construct a packed cache line starting from any logical linear address using Alg.~\ref{alg:data-transformation}. The transformation is semantically powerful because $F$ is not bound to a fixed structural template and, in principle, may be chosen from a broad class of functions, subject to hardware realizability.


\begin{figure*}[t]  
  \centering
  \includegraphics[width=0.95\textwidth]{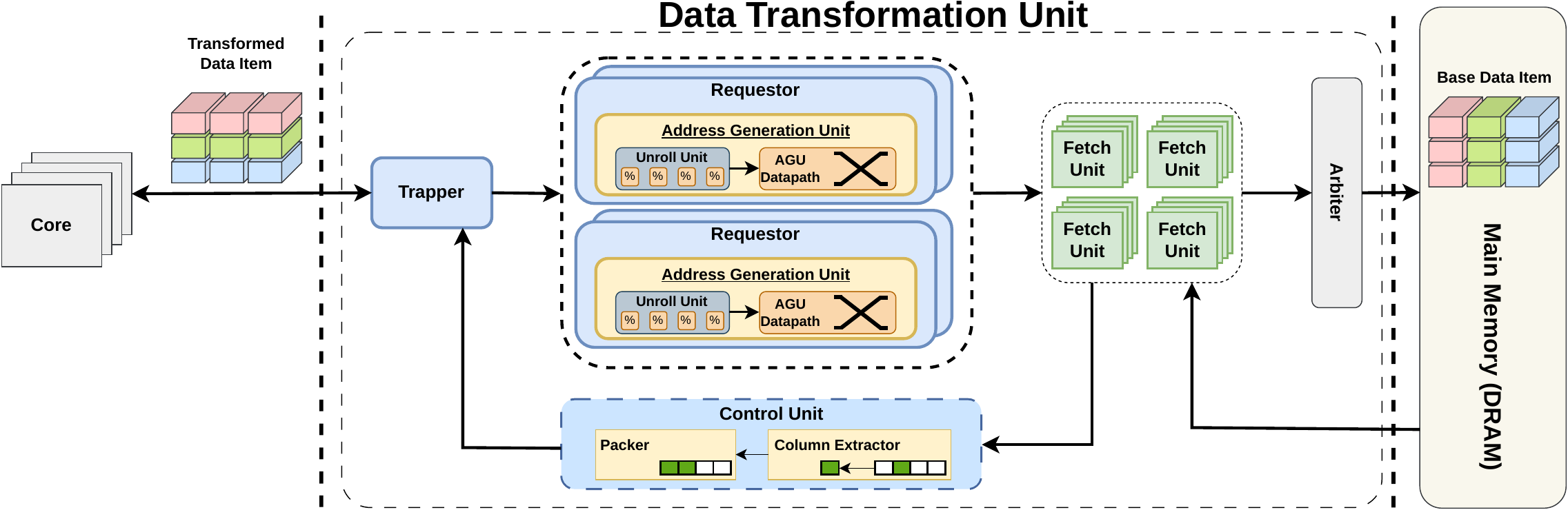}
  \caption{Bird's eye view of the DTU design. The DTU is instantiated between the processor-side LLC and main memory. It traps cache line misses for a set of registered transformed data objects. The DTU dynamically constructs the content of cache lines to be returned. This is done by applying the corresponding DTL kernel to retrieve contiguous logical data items that are scattered in positional space.}
  \label{fig:dtu-abstract}
\end{figure*}

\section{System Overview}\label{sec:sys-overview}
We now introduce the architectural support and programming infrastructure for hardware-assisted data transformation. The proposed system consists of two main elements, namely the Data Transformation Unit (DTU) and the associated Data Transformation Language (DTL). Together, they enable programmable on-the-fly data transformation, avoiding the need to materialize transformed layouts in memory. The DTU is logically instantiated between the LLC and main memory. It is programmed using layout transformation kernels written using the provided DTL and compiled using the associated DTL compiler.



\subsection{Data Transformation Unit (DTU)}
The DTU is a hardware module that exposes a programmable interface for providing transformed views of data objects. Figure~\ref{fig:dtu-abstract} provides an overview of the DTU. The DTU effectively converts a single cache line request into (potentially) multiple main memory accesses for individual data items. It then reassembles the result into a cache line returned to the processor according to a programmable layout transformation kernel. This operation corresponds to the retrieval of a set of data items that are sequential in logical space but scattered in positional space.

As mentioned in Section~\ref{sec:data-transformation}, data transformation can be performed entirely in virtual or physical address space. For simplicity of discussion and implementation, we consider a DTU design operating with physical addresses. As such, virtual-to-physical address translations are performed before requests reach the DTU, and both logical and positional addresses are always physical addresses. The proposed DTU logically sits between the LLC and main memory and traps any transaction that falls within a configurable range of physical addresses, called the \emph{shadow range}. 
The shadow range is chosen to be non-overlapping with the range of physical addresses assigned to main memory or any other memory-mapped device  and is exposed via a TileLink~\cite{tilelink} Manager node~\cite{chipyard}. Normal memory accesses do not fall within the shadow range, are not routed through the DTU, and incur no extra access latency.


Consequently, for requests targeting memory within the shadow range, the DTU sits on the critical path to main memory. This placement allows the DTU to intercept memory requests, issue independent requests to main memory, and reorganize the returned data before it is delivered back to the cache hierarchy. Due to this placement, any additional delay introduced by the DTU directly increases the reorganized data access latency. As a result, the DTU microarchitecture must be designed to offer a compromise between supporting rich and programmable transformations while minimizing reorganization latency.

As shown in Figure ~\ref{fig:dtu-abstract}, the DTU is composed of several pipeline stages that perform specialized functions. A single DTU supports the simultaneous configuration of multiple data transformation kernels. The \textbf{Trapper} module is responsible for receiving LLC misses handled by the DTU and generating compliant responses. The Trapper also matches incoming requests against the set of active configurations to determine how the request should be processed.

Each \textbf{Requestor} module instance handles a single kernel and transforms an incoming request into a sequence of outbound memory accesses. The Requestor effectively recovers logical coordinates from incoming logical addresses and issues the necessary sequence of positional data accesses to construct cache lines of logically contiguous data items. The internal machinery responsible for this process is the \textbf{Address Generation Unit} (AGU), which contains two components: the \textbf{Unroll Unit} and the \textbf{AGU Datapath}. The Requestor provides the AGU with incoming requests and associated metadata, while the AGU generates the memory addresses used for outbound memory accesses.

Upon receiving a request, the AGU first passes the address to the Unroll Unit, which evaluates $L^{-1}$ (Eq.~\eqref{eq:rev-linear-index}) to recover the logical coordinate vector. These logical coordinates are then written into registers within the AGU Datapath. The coordinates are subsequently dispatched for computation, after which the coordinate successor algorithm is applied (Algorithms~\ref{alg:next-coord} and~\ref{alg:data-transformation}). Specifically, the AGU Datapath evaluates the compiler-generated function $F$ (Eq.~\eqref{eq:F}) for each coordinate vector until enough addresses have been generated to construct a full cache line.

The Requestor receives each generated address, converts it into a bus-aligned memory request, attaches metadata (e.g., cache line placement and valid-data mask), and forwards the request to the \textbf{Fetch Unit}. The Fetch Unit forwards requests to main memory and returns responses to the \textbf{Control Unit}. Within the Control Unit, the \textbf{Column Extractor} extracts the relevant data from each response and forwards it to the \textbf{Packer}. The Packer uses the metadata generated by the Requestor to place each extracted element in its correct position within the cache line being constructed. Once a full cache line has been assembled, the Control Unit forwards it to the Trapper, which returns the response upstream to the cache hierarchy.

Nearly every component of the DTU can be parameterized, allowing the design to be instantiated with different performance characteristics. The parameters include the supported configurations, pipeline width, memory-level parallelism, and AGU characteristics (i.e., the set of supported $F$ and $L$).

\subsection{Data Transformation Language (DTL)}\label{sec:dtl-overview}

The DTL is a small domain specific language (DSL) for concisely defining layout transformation kernels. Due to space constraints, we provide a full formal DTL specification in the appendix in Listing~\ref{lst:appendix-grammar} and use this section to describe its key features. Our DTL allows the programmer to provide a concrete instantiation of $L$, $L^{-1}$, $N$, and $F$---see Section \ref{sec:data-transformation}. These functions are implicitly specified by the structure of a DTL kernel, which closely resembles C code to remain friendly to end-users.

\lstset{
    language=C,               
    basicstyle=\ttfamily\tiny, 
    keywordstyle=\color{blue},
    commentstyle=\color{gray},
    stringstyle=\color{red},
    numbers=left,             
    numberstyle=\tiny\color{gray},
    stepnumber=1,
    showstringspaces=false,
    breaklines=true,
    frame=single,
    escapeinside={(*@}{@*)},
    alsoletter={?},
    emph=[1]{out},emphstyle=[1]{\color{red}\bfseries},
    emph=[2]{isEven?,switch,pad?,isEdge?},emphstyle=[2]{\color{purple}},
}

\newcommand{\outst}{{\color{red}\textbf{\texttt{out}}}\xspace}
\newcommand{\forst}{{\color{blue}\texttt{for}}\xspace}
\newcommand{\isevenst}{{\color{purple}\texttt{isEven?}}\xspace}
\newcommand{\switchst}{{\color{purple}\texttt{switch}}\xspace}
\newcommand{\padst}{{\color{purple}\texttt{pad?}}\xspace}
\newcommand{\isedgest}{{\color{purple}\texttt{isEdge?}}\xspace}

Consider once again the simple example of a matrix \matr{B} positionally stored in row-major order and logically accessed as its transpose $\matr{B}^\intercal$. 
Listing~\ref{lst:example-dtl-kernel} shows an example of a DTL kernel that exports the logical view of $\matr{B}^\intercal$, assuming that \matr{B} is a $16\times16$ matrix.

\begin{minipage}{0.95\linewidth}
\begin{lstlisting}[caption={Example DTL kernel for implicitly transposing the 16x16 matrix \matr{B} from Section~\ref{sec:motivating-example}.},label={lst:example-dtl-kernel}]
int ROWS = 16; (*@\label{line:dtl1_const}@*)
int COLS = 16; (*@\label{line:dtl1_const}@*)

for (int col = 0; col < COLS; col++) (*@\label{line:dtl1_for1}@*) {
    for (int row = 0; row < ROWS; row++) (*@\label{line:dtl1_for2}@*) {
        out = row*COLS+col; (*@\label{line:dtl1_out}@*) } }
\end{lstlisting}
\end{minipage}

From the above example and with reference to the full specification, we note a few distinctive structural features of a DTL kernel. First, a set of integer constants can be declared (Line~\ref{line:dtl1_const}). Second, the main body of the kernel must follow a set of nested \forst loops (Lines~\ref{line:dtl1_for1} and~\ref{line:dtl1_for2}). Each \forst loop must declare and initialize a fresh loop variable, provide a loop condition using the less-than "<" operator against an explicit or previously-defined constant, and include a plus-one increment rule for the current loop variable. The innermost loop must contain one or more \outst statements (Line~\ref{line:dtl1_out}). The right-hand side of the \outst statement calculates the positional address at the current iteration. Furthermore, the innermost loop body contains an optional conditional (one of $>, \geq, <, \leq, \isevenst, \switchst, \padst$, or $\isedgest$) followed by one or more \outst statements. Parameters to conditional statements can only be loop variable identifiers. Each conditional branch must contain the same number of out statements such that $L$ and $L^{-1}$ are efficiently computable. Additionally, a limited number of fixed-size data arrays can be indexed within an out-statement, and multiple out-statements in the innermost loop can be used, effectively defining an additional dimension whose bounds equal the number of out-statements and, therefore, extend the logical coordinate space accordingly. These capabilities further enhance the generality of DTL, enabling a single kernel to represent a much broader class of transformations.




Despite its resemblance to imperative C-like syntax, a DTL kernel does not execute iteratively in the conventional sense. Rather, it defines a mapping $F$ from the logical coordinate space (the Cartesian product of the loop index domains) to offsets within a data object in positional space. The logical coordinate space defines a multi-dimensional data object, where each loop corresponds to a dimension, the bounds define that dimension’s extent, and the order of loop nesting determines the linearization order. From this dimensional definition, an access stride structure and thus a concrete instantiation of $L$ and $L^{-1}$ are derived, while the traversal order over the logical coordinate space defines the coordinate successor function $N$. 

The \outst statement in the innermost loop is the concrete realization of $F$: it computes an offset address as a function of constants and loop indices, identifying the data element to be accessed in the positional layout. In our current implementation and targeted benchmarks, we support the following set of operators within $F$: $\{+,-,*,nop\}$. These elements are then exposed in the traversal order of the logical coordinate space, thereby defining the linearized layout of the transformed data object. Intuitively, the DTL kernel exposes a transformed logical view of the base data object, as if C code of similar structure was actually executed to materialize the transformation.

Listing \ref{lst:rubix-cube-dtlkernel} showcases a more complex DTL kernel that takes slices along the depth of a 3D tensor and alternately rotates each 2D slice left or right. DTL supports a restricted class of conditionals that allow the mapping function to vary as a function of the coordinate vector. To do so, the kernel uses the conditional operator \isevenst which allows construction of piecewise mappings over the logical coordinate space. These predicates are evaluated using bounded combinational logic and add no additional pipeline latency. 

\begin{minipage}{0.95\linewidth}
\begin{lstlisting}[caption={Example DTL kernel applying alternate left/right rotations to each slice of a 3D tensor.}, label={lst:rubix-cube-dtlkernel}]
int stride_d = 400*400;
int stride_h = 400;
int MAX_DEPTH = 100;
int MAX_HEIGHT = 400;
int MAX_WIDTH = 400;

for (int d = 0; d < MAX_DEPTH; d++) {
  for (int w = 0; w < MAX_WIDTH; w++) {
    for (int h = 0; h < MAX_HEIGHT; h++) {        
      if (isEven? d)  // Rotate Right
        out = d*stride_d+(MAX_HEIGHT-1-h)*stride_h+w;
      else            // Rotate Left
        out = d*stride_d+h*stride_h+(MAX_WIDTH-1-w); } } }
\end{lstlisting}
\end{minipage}



\subsection{DTU Runtime and API}\label{sec:dtu-runtime} 

In addition to the DTU hardware and the DTL compiler, we also provide the necessary runtime drivers and programming APIs to leverage DTU functionalities in user applications. These components include (1) DTU kernel compilation and registration, (2) memory management and access methods, and (3) synchronization primitives. The runtime is implemented in part as an \textbf{OS DTU driver} and in part as a \textbf{DTU user-space application library}. 

\begin{figure}[htp]
  \centering
  \includegraphics[width=0.48\textwidth]{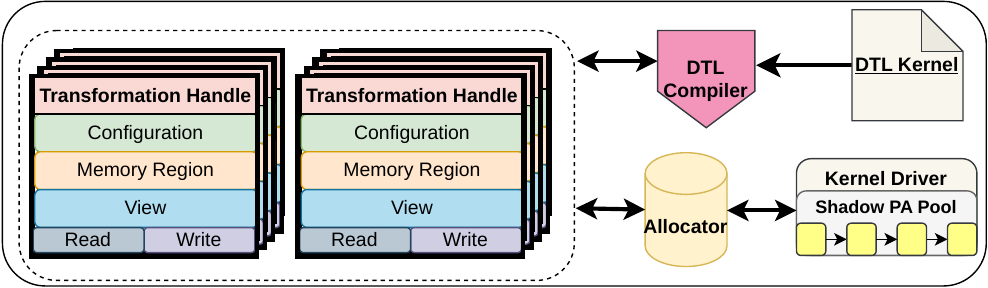}
  \caption{DTU runtime overview: transformation handles encapsulate the resources necessary for DTL kernel deployment.}
  \label{fig:dtu-runtime}
\end{figure}

At system initialization, the OS driver allocates a contiguous range of physical addresses backed by main memory that will be used to store base objects. It also allocates the shadow range from where physical addresses for non-materialized transformed objects are allocated. The driver also instantiates a memory manager that will handle memory allocations for both base objects and transformed logical objects. Finally, the driver instantiates low-level bidirectional communication interfaces with user-space that are used by the DTU library. The primary abstraction exported to user applications by the DTU library is the \emph{transformation handle}. The transformation handle is an object that encapsulates all the methods and resources required to register and use a DTU kernel, and to interact with logical objects. A high-level view of this interface is shown in Figure~\ref{fig:dtu-runtime}.

The first step for an application is to allocate enough memory, through the library-exported API, to store the base object in positional space. 
A successful first step returns a valid transformation handle to the application. Upon success, the following two resources are set up: (1) \emph{Memory Region:} a continuous memory-backed range of physical addresses for the base object; (2) \emph{Write Port:} a virtual-to-physical mapping to this range is established. The application can use this write pointer to initialize/update the base object. 
Writes to the base object use non-temporal (NT) stores targeting the DRAM-backed region, ensuring that updates bypass the cache hierarchy and are immediately visible to the DTU. 

Second, the application registers a compiled DTU kernel using the previously obtained handle. 
Upon successful registration, the size of the logical object is computed and the following is set up: (1) \emph{Configuration:} a Requestor module inside the DTU is configured with the provided kernel and the resulting module configuration is reflected in the handle; (2) \emph{View:} a continuous range of shadow physical addresses for the transformed object; (3) \emph{Read Port:} a virtual-to-physical mapping such that a read pointer can be used by the application to access the transformed logical data object. The read pointer can then be used to access the logical view of the object, 
and can be passed in to unmodified compute kernels expecting the logical layout. 

\setcounter{listing}{\value{lstlisting}}
\begin{listing}[htp]
\centering

\begin{tcolorbox}[
  enhanced,
  colback=codebg,
  boxrule=0pt,
  breakable
]
\input{_minted/vs.style.minted}

\begingroup
\fvset{
  fontsize=\scriptsize,
  numbers=left,
  numbersep=6pt,
  breaklines=true,
  tabsize=2
}
\input{_minted/3552775C3B1A2B5A8FBFD9E396C33FA0.highlight.minted}
\endgroup
\end{tcolorbox}

\caption{DTU runtime and API workflow for the matrix-multiplication example
introduced in Section~\ref{sec:motivating-example}. The code initializes the positional
representation of \matr{B}
and accesses the resulting $\matr{B}^\intercal$ logical view using an otherwise
unmodified matrix-multiplication kernel. 
}
\label{lst:dtu-programming-interface}
\end{listing}

Listing~\ref{lst:dtu-programming-interface} shows an example matrix-multiplication workflow using the DTU to perform an on-the-fly matrix transpose. First, the application allocates positional memory space for matrix $\matr{B}$ (Line 3) and initializes the matrix (Line 8). Next, the transpose DTL kernel (Listing~\ref{lst:example-dtl-kernel}) is compiled (Line 11) and programmed into the DTU (Line 15). Finally, the application obtains a pointer to the transformed $\matr{B}^{\intercal}$ in logical address space (Line 18), which is passed directly to the matrix-multiplication kernel (Line 21).



\subsection{Coherence and Synchronization}

Note that all read accesses to the transformed object are internally issued through shadow physical address (SPA) regions, which lie outside the physical address range of main memory and provide a level of indirection over the underlying in-memory object that enables address translation by the DTU. However, this indirection alone does not guarantee data freshness. For example, a transformed read may bring a cache line into the cache hierarchy, after which the base object in main memory is updated, leaving the cached line stale. To ensure that subsequent reads observe updated data, we employ \textit{SPA remapping}. Specifically, when visibility of updates is required, the runtime allocates a new SPA region and remaps the virtual addresses to this new region, forcing future accesses to hit a different physical address. Because LLCs are physically tagged, cache lines associated with the previous mapping do not alias with the new one, preventing stale data from being reused without requiring explicit cache invalidation. This mechanism enables multiple simultaneous views of the same data object and provides software control over when updates become visible. Consequently, multiple cores can safely hold different transformed views of the same base object concurrently. 

SPA remapping requires updating page-table entries, invalidating corresponding TLB entries, refetching data from DRAM, and executing a few MMIO writes to the DTU's configuration registers. Although these costs are non-trivial, remapping is seldom required given the discrete read/write phase-based access patterns of our target workloads, allowing these overheads to be well amortized. Furthermore, using hugepages greatly reduces the TLB invalidation overhead.

Lastly, because the shadow PA space is finite, stale cache lines could potentially be encountered if address range reuse occurs due to wraparound. Although extremely unlikely in practice (for instance, in a RISC-V Sv48-based~\cite{riscv-isa-spec} virtual memory system, this range spans from the maximum memory-backed address up to $2^{47}-1$), correctness is guaranteed by issuing a cache flush upon wraparound---a cost that is well amortized due to its rare occurrence.

\section{Implementation}\label{sec:implementation}
This section describes the implementation of the DTU’s Address Generation Unit (AGU) and the DTL compiler.




\subsection{Address Generation Unit (AGU)}\label{sec:agu}

Address generation is fully decoupled from the rest of the system. A well-defined interface between the Requestor and the AGU enables a producer–consumer interaction model. The Requestor activates the AGU by passing in a new request and in turn the AGU will output a sufficient number of addresses to construct a cache line. The AGU has two key components: the \textbf{Unroll Unit} which is the physical realization of $L^{-1}$ (Eq.~\eqref{eq:rev-linear-index}), and the \textbf{AGU Datapath} which performs the computation defined by $F$ (Eq.~\eqref{eq:F}). Putting this together, the Unroll Unit recovers (via $L^{-1}$) the coordinate vector from an input linear logical address ($L$), which is used to initialize the AGU Datapath; the datapath then iteratively applies $F$
and the coordinate successor function $N$ to generate the sequence of addresses required to construct a cache line.

\paragraph{Unroll Unit}\label{sec:unroll-unit}
The Unroll Unit transforms a logical linear address into its corresponding coordinate vector, providing a programmable realization of Eq.~\eqref{eq:rev-linear-index}. Computation of $L^{-1}$ requires a series of integer divisions, which have notoriously long multi-cycle latency in their hardware implementation~\cite{hardware-division}. Because these divisions must be performed sequentially on the critical path to main memory, these multi-cycle divisions could easily dominate any performance gains made by improved cache locality. To combat this, we use the insight that every divisor is known at compile time to perform each operation using the \emph{magic division algorithm}~\cite{hackers-delight2e}. The magic division algorithm allows an integer division operation to be performed using a precomputed set of ``magic'' numbers and multiplication, addition, and bit shift operations instead of costly hardware divisions. This algorithm allows each division to be performed in two deterministic cycles, avoiding a complex long or variable latency implementation while maintaining a high operating frequency. 

Given a DTL kernel, the compiler (Section~\ref{sec:dtl-overview}) extracts stride information from the loop nest that defines the linearization order and uses it to compute the corresponding magic constants. These constants parameterize the divisions in Eq.~\eqref{eq:rev-linear-index} and are written to configuration registers within the Unroll Unit.

The Unroll Unit can be flexibly instantiated to support any number of maximum dimensions or offset widths. Additionally, the DTU allows for runtime configuration with different coordinate dimensions. Accordingly, the Unroll Unit intelligently utilizes dimensional knowledge of the data object to only perform the needed number of divisions - avoiding incurring unnecessary latency. Once coordinates are recovered, they are passed to the next stage of the pipeline: the AGU Datapath.

\paragraph{AGU Datapath}\label{sec:agu-datapath}

\begin{figure}[t]
  \centering
  \includegraphics[width=0.48\textwidth]{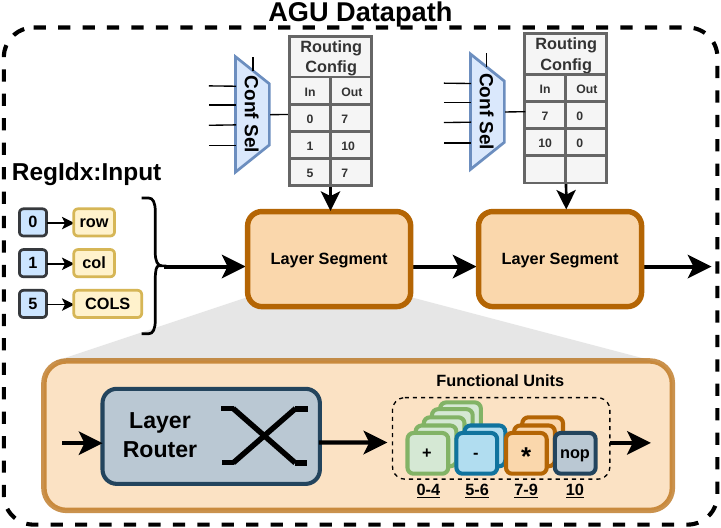}
  \caption{AGU datapath configured to compute the $F(row,col)=row*COLS+col$ mapping given by the matrix transpose DTL kernel of Listing~\ref{lst:example-dtl-kernel}.}
  \label{fig:agu-datapath}
\end{figure}

The AGU Datapath takes a logical coordinate vector as input and produces the sequence of addresses corresponding to the data items that compose a cache line. It provides a programmable realization of the functions $F$ and $N$.
Function $F$ maps each coordinate vector to its corresponding positional address, while $N$ defines the progression to the next coordinate. The ordering induced by $N$ matches the iteration order of the nested loop structure in a DTL kernel.

Operationally, each coordinate vector is fed into a pipelined computation that evaluates $F$ to produce an address. The coordinate is then updated according to $N$, and this process repeats until a sufficient number of addresses have been generated to construct a cache line.

The AGU Datapath is organized as a sequence of pipeline stages composed of simple functional units. In our implementation, these units are $\{+, -, *, \text{nop}\}$, matching the operator set currently supported by DTL. Each stage can be instantiated with a configurable composition of functional units, while programmable routing modules between stages map inputs to the appropriate units to realize $F$. A high level view of the AGU Datapath is shown in Figure \ref{fig:agu-datapath}.

This layered organization enables tight pipelining across successive coordinate vectors, allowing each computation to overlap with its predecessor and amortize pipeline latency. The routing configuration is a simple input-to-output mapping, and serves as the compiler target for a DTL kernel.

\subsection{DTL Compiler} \label{subsec:dtl-compiler}
Each DTL kernel is compiled into a layout transform configuration, which includes the parameterization of the Unroll Unit, the coordinate successor function $N$, and the routing configuration for the AGU Datapath. Configurations can be compiled online or offline and can be dynamically attached, removed, or saved at runtime. The extraction of the Unroll Unit configuration and coordinate successor function from a kernel was described in the previous section; the routing configuration, which realizes $F$, is derived from the kernel's \outst-statements. Each \outst-statement specifies a sequence of operators and operands, with operands mapped to hardware locations and operators mapped to functional units in the AGU Datapath. The dependency structure of these expressions forms an abstract syntax tree, where leaves correspond to inputs and higher nodes depend on the results of lower nodes. Operations at the same tree level can be scheduled onto the same layer of the AGU Datapath, allowing the compiler to greedily construct a VLIW-like schedule to maximize per-cycle throughput and minimize pipeline depth. Each functional unit has a layer-specific index, and the compiler realizes a schedule by defining an input index to output index mapping for each router module between layers. The compiler is aware of the current DTU parameterization, and attempts to realize a schedule that fits the current parameterization of functional units and layer count by moving operations across layers when legal. Our compiler implementation also supports a number of standard optimization passes (i.e., constant propagation, constant folding, dead code elimination, value numbering, copy propagation) for reducing resource count and increasing schedule quality. An example schedule for the running matrix transpose example is shown in Figure~\ref{fig:agu-datapath}.

The primary architectural constraint is that the DTU lies on the critical path to memory; consequently, any latency introduced within the DTU directly increases memory access latency. Minimizing this internal latency therefore drives both the hardware and compiler design. To address this, the system enables flexible computation schedules that execute multiple operations per cycle, while tightly pipelining successive computations of $F$. This approach significantly expands the space of profitably realizable layout transformations, providing a highly general and efficient framework for mapping complex transformations onto the hardware.

\subsection{DTU Access Latency}
Normal memory accesses bypass the DTU logic entirely. Therefore, normal memory accesses experience no additional latency. The shadow range is exposed via a TileLink Manager node, and only memory accesses targeting this range will be routed to the DTU.

Accesses to DTU-transformed objects are routed through the DTU and therefore incur additional latency relative to ordinary memory accesses. This latency is not constant: producing a single logical cache line may require fetching data from multiple, potentially disparate, positional locations in DRAM. The total latency thus depends on both the specific transformation being performed and the number and location of the underlying memory accesses required to construct the response.

The structural latency of the DTU pipeline itself—including address generation, routing, and index-calculation logic—ranges from 13 to 27 cycles depending on the active DTL kernel. This variation stems from the variable number of integer divisions required to compute $L^{-1}$ in the Unroll Unit. Even at the high end of this range, the structural latency overhead is small relative to a conventional DRAM access, which typically requires hundreds of cycles. Moreover, prior work has shown that prefetching can effectively hide much of the additional latency introduced by on-the-fly transformation~\cite{strickler2025adms-rme}.

\section{Evaluation}\label{sec:evaluation}
In this section, we evaluate the DTU across a range of workloads and configurations to quantify its impact on execution time, memory system behavior, and hardware cost.

\subsection{System Setup}
We use FireSim~\cite{karandikar2018firesim} as our simulation environment. FireSim
is an FPGA-accelerated, cycle-exact, full-system simulator.
It enables simulation of a complete SoC, with the FPGA
operating at 100 MHz while the simulated target runs at a
configurable frequency. Simulation results are not affected
by the FPGA speed limitation, as a specialized compiler
infrastructure decouples the target design from the FPGA
host~\cite{golden-gate}, enabling realistic performance evaluation. On FireSim, we instantiate a quad-core Berkeley-Out-Of-Order Machine (BOOM)~\cite{asanovi2015boom} RISC-V SoC using Chipyard~\cite{chipyard} and implement the proposed DTU hardware block, written in Chisel HDL~\cite{chisel2012}, within the SoC, operating in the memory bus clock domain at 1GHz. We use Linux kernel version 6.2 and a Buildroot-based root filesystem~\cite{buildroot} as the runtime environment. The details of our platform are shown in Table~\ref{tab:system-setup}.



\begin{table}[th]
    \caption{Evaluation platform specifications.}
    \label{tab:system-setup}
    \vspace{-0.1in}
    \centering
    \small
    \begin{tabular}{cl}
        \toprule
        \textbf{Component} & \textbf{Specifications} \\
        \midrule
        \textbf{CPU} &\makecell[l]{4x BOOM@3.2GHz, Out-Of-Order, \\ 
        3-wide, ROB: 96, LSQ: 24/24 \\
        L1: 32K(I)/32K(D) (8-Way), 6 MSHRs} \\
        \textbf{Prefetcher} & \makecell[l]{L1 4-Ahead Multi Next Line~\cite{bar-fetchers}} \\
        \textbf{Shared L2 Cache} & \makecell[l]{1MB (16-way), 1 Bank~\cite{inclusive-L2}} \\
        \textbf{DTU} & \makecell[l]{16 Fetch Units, 8 Requestors, 1GHz} \\
        \textbf{Main Memory} & \makecell[l]{4GB DDR3 1066 MHz, 1 Rank, 8 Banks} \\
        \hline
        \textbf{Operating System} & Buildroot Linux v6.2.0\\
        \bottomrule
    \end{tabular}
\end{table}

\subsection{Benchmarks}\label{}
For benchmarks, we use the following five representative workloads: 

\textbf{Database Column Projection} refers to a database query operation of the form \texttt{SELECT C1, C2, \dots CN FROM TABLE}, which is commonly used in analytical processing. 
The performance of this query is highly sensitive to whether the database uses column- or row-wise storage and the number of projected columns.
In this benchmark, we run this query on an in-memory database using column-wise, row-wise storage, or DTU over a row-wise storage backend. The goal is to show that starting with a row-store backend, DTU allows us to always unlock the benefits of a column-store~\cite{abadi2008column}.
Note that the DTU can enable the development of 
a hardware-assisted Hybrid Transactional/Analytical Processing \cite{Song2024SurveyHybridTA,Ozcan2017,roozkhosh2023edbt-rme} engine that exhibits the benefits of a row-store for transactions and a column-store for analytical queries. The benchmark we use here is adopted from the RME framework~\cite{roozkhosh2023edbt-rme}. 


\textbf{Im2Col} is an optimization technique used in convolutional neural networks to speed up convolution operations by converting them into standard matrix multiplications. Instead of sliding a filter across an input image using slow, nested loops, Im2Col extracts every overlapping image patch the filter will process, flattens them into individual columns and assembles them into a single large 2D matrix. The convolutional filters are similarly flattened into another matrix. By restructuring the data in this manner, the entire convolution step can be computed simultaneously as a single, optimized General Matrix Multiplication (GEMM).
For this benchmark, we apply Im2Col over a $720\times1080$ RGB image, followed by a GEMM, to execute a convolution operation with one filter.
For the CPU baseline, we compare against the Darknet~\cite{yolo} Im2Col implementation.

\textbf{Image Augmentation} is a preprocessing technique used in training vision networks meant to improve generalization and prevent overfitting. For this benchmark, we perform a fused NHWC to NCHW permutation and augmentation over a batch of images~\cite{cnns_on_cpus19}. We perform augmentation from the set $\{\texttt{RotateLeft},\texttt{RotateRight},\texttt{ReflectX},\texttt{ReflectY},\\\texttt{RotateLeft+\\ReflectX},\texttt{RotateRight+ReflectX},\texttt{ReflectX}\\\texttt{+ReflectY} \}$ producing seven new unique views of each image in the batch. We follow the augmentation with an intensity scaling operation that further doubles the set of augmented images.


\textbf{Tensor Unfolding} is an operation that converts a higher-order tensor into a matrix (matricization) such that one axis becomes a row while the rest become columns. Formally, for an $n$-order tensor $\mathcal{X}\in \mathbb{R}^{I_1\times I_2 \dots\times I_n}$, $I_j\in \mathbb{Z}^+$, a mode-$k$ unfolding yields a matrix $\mathcal{X}_{(k)} \in \mathbb{R}^{(I_k\times\prod_{j \neq k} I_j)}$. Unfolding across each mode yields different memory access characteristics. For this benchmark, we apply an unfolding operation and then perform a Hadamard product using the resulting matrix. We utilize a $32\times32\times32\times32$ tensor. 

\textbf{Vol2Col} is a generalization of Im2Col to four dimensions, where a 4D tensor is lowered into a matrix suitable for applying the 3D convolution kernel via matrix multiplication. In this case, the output matrix scales in size cubically with the kernel size. For this benchmark, we apply Vol2Col to an input tensor and then perform the GEMM convolution. We fix the volume size to $3\times16\times128\times128$ and the number of output channels to one. We utilize the PyTorch Vol2Col implementation~\cite{pytorch-vol2col}.

\vspace{-0.1in}

\subsection{Effect on Execution Time and WSS}

In this subsection, we evaluate DTU's effects on performance and memory consumption. 

\begin{figure}[htp]

  \centering
  \begin{subfigure}[b]{\columnwidth}
    \centering
    \includegraphics[width=\linewidth]{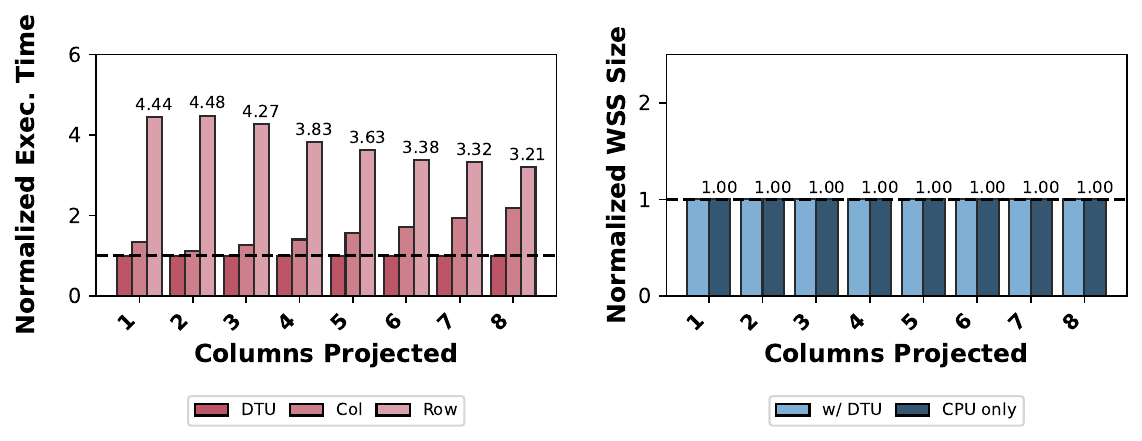}
    \vspace{-0.2in}
    \caption{Column Projection}
    \label{fig:dbproj-boom}
  \end{subfigure}


  \begin{subfigure}[b]{\columnwidth}
    \centering
    \includegraphics[width=\linewidth]{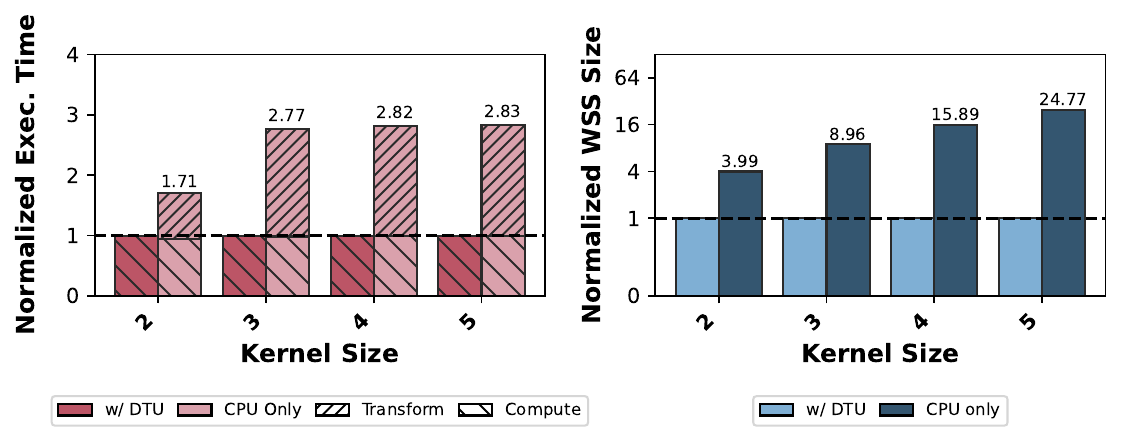}
    \vspace{-0.2in}
    \caption{Im2Col}
    \label{fig:im2col-boom}
  \end{subfigure}


  \begin{subfigure}[b]{\columnwidth}
    \centering
    \includegraphics[width=\linewidth]{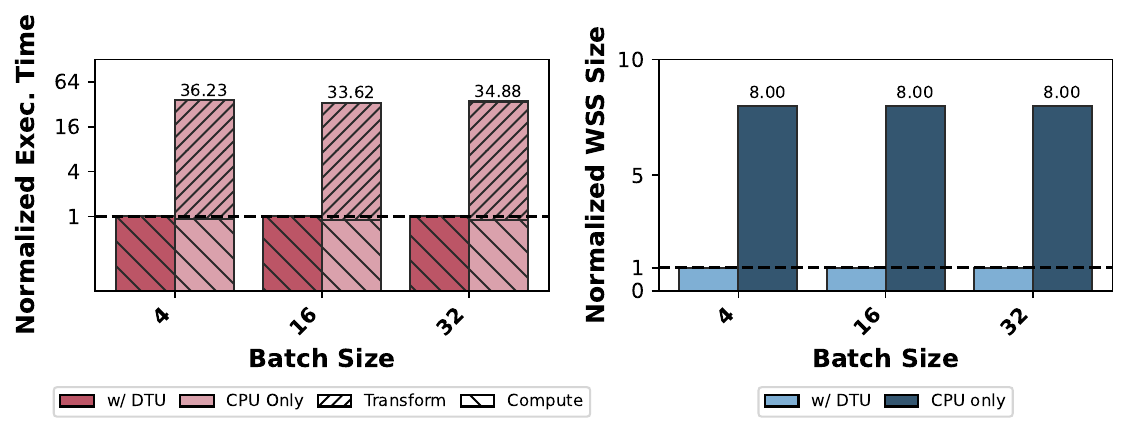}
    \vspace{-0.2in}
    \caption{Image Augmentation}
    \label{fig:imgaug-boom}
  \end{subfigure}


    \begin{subfigure}[b]{\columnwidth}
    \centering
    \includegraphics[width=\linewidth]{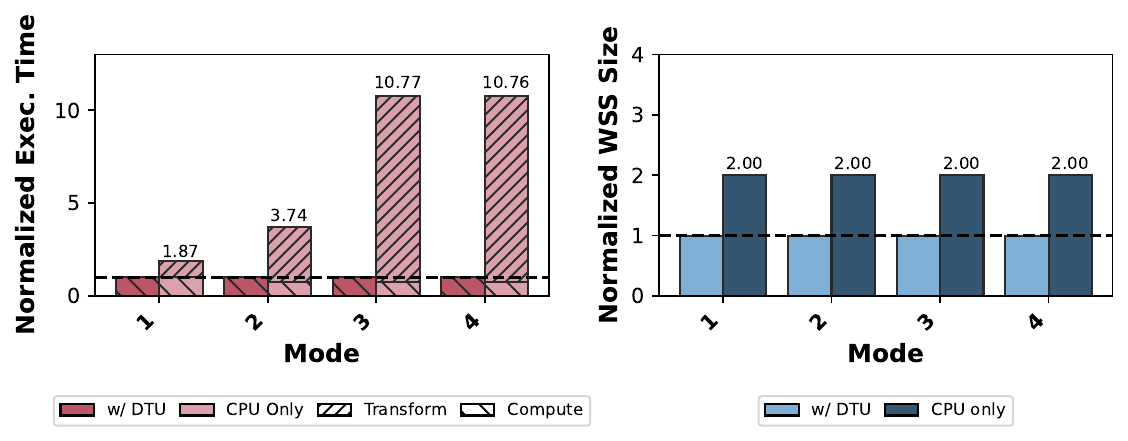}
    \vspace{-0.2in}
    \caption{Tensor Unfolding}
    \label{fig:tensor-unfold-boom}
  \end{subfigure}


  \begin{subfigure}[b]{\columnwidth}
    \centering
    \includegraphics[width=\linewidth]{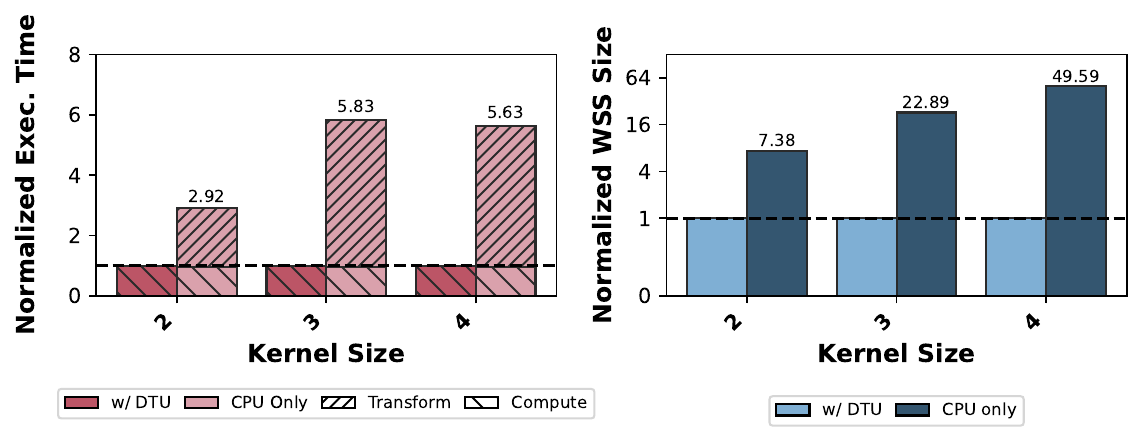}
    \vspace{-0.2in}
    \caption{Vol2Col}
    \label{fig:vol2col-boom}
  \end{subfigure}

  \caption{Effects on execution time and WSS.}
  \label{fig:microbench-grid}
\end{figure}

Figure~\ref{fig:dbproj-boom} shows the normalized execution time results for the \emph{Database Column Projection} benchmark. In a row-wise layout (Row), only a fraction of each fetched cache line contains useful data, as unwanted attributes are interleaved with the target fields, resulting in suboptimal performance. In contrast, a columnar layout (Col) stores each attribute contiguously, allowing accesses to retrieve only the relevant data and thereby improving spatial locality and reducing unnecessary memory traffic, resulting in measurably better performance.

The DTU, on the other hand, provides a contiguous view of the selected attributes even when the underlying table is stored in a row-wise layout, outperforming both Row and Col. While both the DTU and a columnar layout achieve ideal spatial locality for individual columns, the DTU outperforms the columnar approach when projecting many attributes. This advantage arises because the DTU interleaves the selected columns into a single contiguous stream, enabling even simple prefetchers to be effective~\cite{strickler2025adms-rme}. In contrast, the columnar layout generates one memory stream per projected column; as the number of projected columns increases, this can exceed the stream-tracking capacity of the hardware prefetcher, reducing its effectiveness.

Figures~\ref{fig:im2col-boom} through \ref{fig:vol2col-boom} show the impact on execution time and working-set size (WSS) of employing our DTU in the four data layout transformation benchmarks. Here, we break down the execution time between the data layout transformation (\emph{Transform}) and subsequent computation (\emph{Compute}). However, in the DTU experiments, the reported execution time only reflects pure computation because the transformation is performed by the DTU hardware on the fly. Comparing the \emph{Compute} portions therefore isolates any runtime overhead incurred when consuming a DTU-provided layout from the performance gains obtained by eliminating explicit transformation.

Note that \emph{Im2Col}, \emph{Image Augmentation}, \emph{Tensor Unfolding}, and \emph{Vol2Col} are all operations that require materialization into a new view, which in the CPU baseline results in substantial data movement and inflates the WSS. The DTU is able to completely eliminate the materialization operation, removing the need for intermediate allocations, drastically reducing WSS. In \emph{Im2Col} and \emph{Vol2Col} in particular, the DTU effectively provides a transformed view that is larger than the base data object---up to $\sim50\times$ in Vol2Col with kernel size four---without requiring explicit materialization in memory. In these benchmarks, it is important to note that the execution time reduction achieved by the DTU primarily comes from the elimination of the explicit data layout transformation cost, while the actual computation cost remains largely unchanged (as low as 1.5\% overhead for Im2Col).

\subsection{Effect on Memory Hierarchy}

We evaluate the impact that the DTU has on the memory hierarchy in terms of cache refills, write-backs, and main memory bandwidth usage.

\begin{figure}[tp]
  \centering

  \begin{subfigure}[b]{0.47\textwidth}
    \centering
    \includegraphics[width=\linewidth]{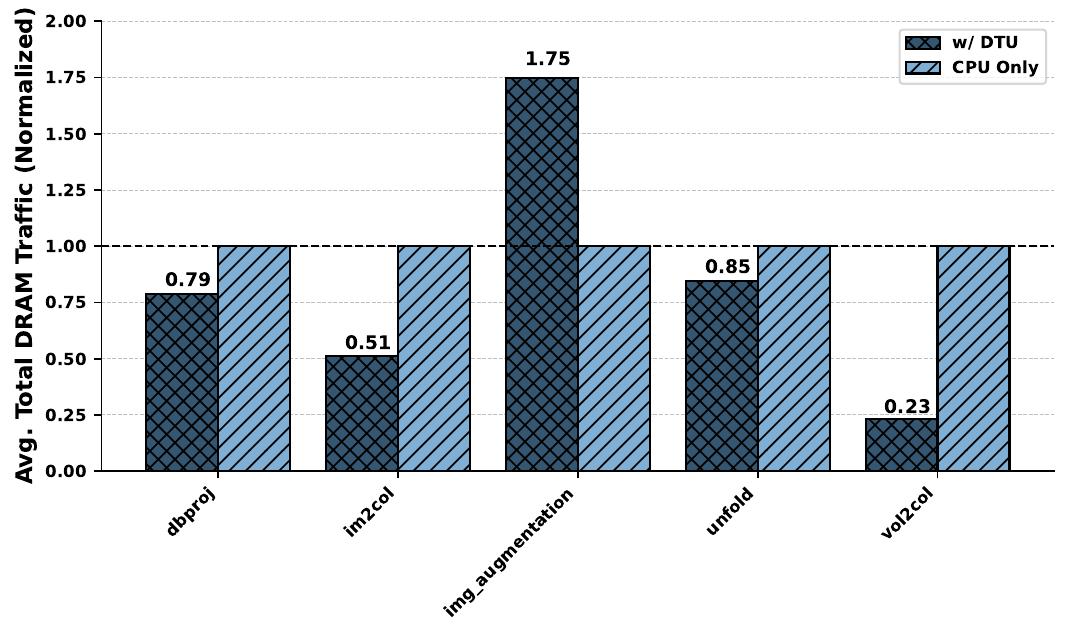}
    \vspace{-0.2in}
    \caption{Average DRAM Traffic}
    \label{fig:dram-traffic}
  \end{subfigure}
  \hfill
  \begin{subfigure}[b]{0.47\textwidth}
    \centering
    \includegraphics[width=\linewidth]{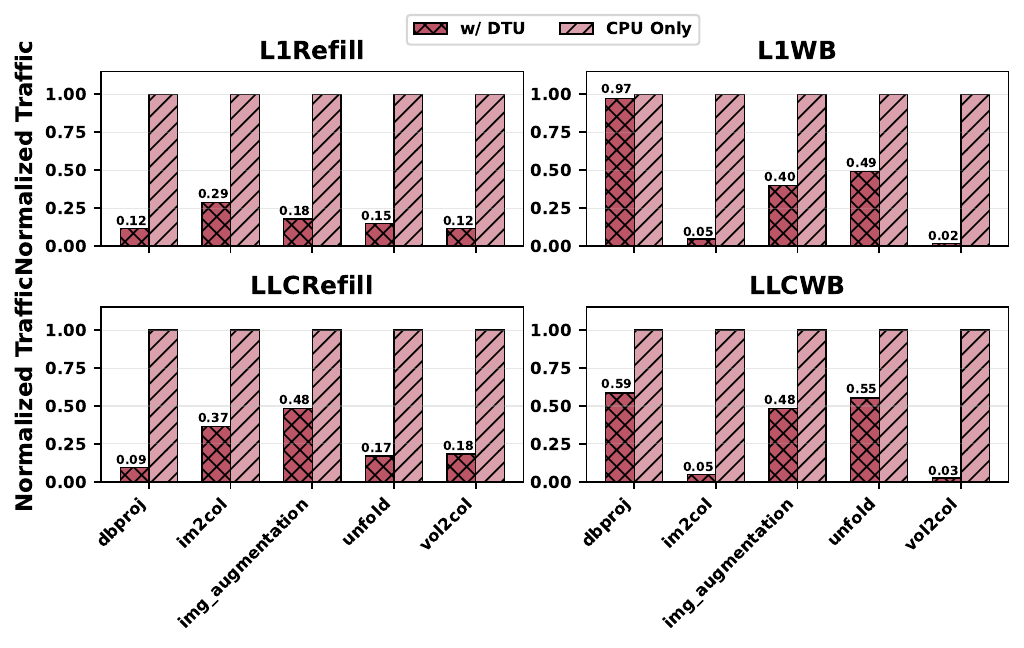}
    \vspace{-0.2in}
    \caption{Average Cache Traffic}
    \label{fig:cache_traffic}
  \end{subfigure}
  \caption{Effects on the memory hierarchy.}
  \label{fig:traffic-analysis}
   \vspace{-0.2in}
\end{figure}

Figure~\ref{fig:traffic-analysis} shows (a) the average usage of main memory (DRAM) bandwidth and (b) the cache statistics for L1 and LLC, with and without using DTU. Here, the results of each benchmark are averaged across all evaluated parameter configurations (referring to the $x$-axis in Figure~\ref{fig:microbench-grid}) to display a single data point for each benchmark. For \emph{Database Column Projection (dbproj)}, we use the \emph{Row} configuration in Figure~\ref{fig:dbproj-boom} as the ``CPU Only'' baseline.


The DTU's impact on DRAM traffic varies by benchmark; while the \emph{img\_aug} benchmark exhibits increased traffic, all of the others show a significant reduction. Increased DRAM traffic occurs in the \emph{img\_aug} case because the DTU may generate multiple DRAM requests for elements that already lie within a single cache-line sized transfer unit. This effect is not fundamental, and could be mitigated with more aggressive request coalescing between DTU-generated addresses, or via reuse of fetched data across related accesses. 


 
Importantly, the DTU consistently reduces both cache misses and write-back traffic (at L1 and LLC) across all benchmarks. This improvement stems from presenting the application logic with densely packed cache lines that exhibit ideal data locality and eliminate intermediate materialization, thus increasing cache efficiency and overall performance.

\subsection{Implementation Overhead}\label{sec:timing-area}
In this subsection, we report the hardware implementation overhead of the DTU.
To quantify the implementation cost of our design, we synthesize a quad-core BOOM SoC 
using Cadence Genus with Hammer automation~\cite{liew22hammer}, targeting the ASAP 7nm~\cite{clark16asap7} technology node.
We measure area and timing overheads for a variable number of parallel DTU configurations (Requestors). 

\begin{table}[th]
\centering
\caption{Area overhead of various DTU configuration counts relative to the quad-core SoC.}
\label{tab:timing-area}
\small
\begin{tabular}{ccccc}
\toprule
\textbf{Configurations} & \textbf{1} & \textbf{2} & \textbf{4} & \textbf{8} \\
\midrule
\textbf{Area (\% of SoC)} & 0.45\% & 0.85\% & 1.47\% & 2.79\% \\
\bottomrule
\end{tabular}
\end{table}

Table~\ref{tab:timing-area} shows the area overhead for four DTU variants. As the configuration count increases, the area overhead ranges from 0.45\% for a single context to 2.79\% for a DTU provisioned with 8 parallel configurations. The 8-configuration variant was used for the results presented in our evaluation. We include smaller configuration counts to show that the DTU can readily be parameterized for different deployment targets and workload needs. All but one of our evaluated workloads utilize only one configuration. The exception is image augmentation which uses seven configurations to export seven different transformed views simultaneously. The configuration count does not affect performance, but restricts the number of transformation views that can be simultaneously exported.
Post-synthesis timing analysis reveals that the DTU meets timing at frequencies up to 1.11~GHz for all configuration counts.

\section{Related Work}\label{sec:related-work}
Several others have explored dynamic data reorganization within the memory hierarchy to better match application access patterns. Early efforts such as GS-DRAM~\cite{gs-dram} and Impulse~\cite{impulse-mc} focused on supporting gather/scatter-style accesses at the memory controller level. The RME~\cite{roozkhosh2023edbt-rme,strickler2025adms-rme,rel-fabric} applied this idea to database systems, while the  TME~\cite{tme-paper} focused on dense line packing for tensor objects. PLIM~\cite{roozkohsh-plim} and CAESAR~\cite{Roozkhosh2022CAESAR} investigate systems level integration of these systems into SoCs that include a tightly coupled FPGA. These prior methods were structurally restrictive in the range of expressible access patterns and constrained to a particular application class. In contrast, our approach puts forward a framework for general cache line construction not tied to a specific domain. 

Systems such as t\"ak\=o~\cite{Schwedock2022tako} and Leviathan~\cite{schwedock-leviathian} extend the cache hierarchy with programmable, near-data execution mechanisms that can react to memory events. In contrast, our approach exposes locality and flexible data views through compiler-directed mappings, rather than relying on programmable execution within the memory hierarchy. 

LEGO~\cite{tavakkoli2025lego}, Linear Layouts~\cite{zhou2026linearlayoutsrobustcode}, and Axe~\cite{hou2026axesimpleunifiedlayout} introduce abstractions for representing tensor layouts that enable compilers to optimize data movement and mapping onto GPU hardware. NVIDIA's CUTLASS~\cite{cutlass-nvida} provides a library of highly optimized GPU kernels that embody such layout and scheduling strategies. In contrast, our approach is processing-element agnostic, operates at memory-block granularity, and does not require active computation on the processing element to realize data transformations.

\section{Future Work}\label{sec:future-work}

There are many optimizations that deserve to be explored both at the hardware and compiler level. Including even more aggressive compiler optimizations could further decrease needed hardware resources, total cycle latency, and area overhead. The loop based structure of DTL makes affine kernels conducive to analysis with the polyhedral model~\cite{polyhedral-thesis}. Building on this structure, a compiler could identify access patterns expressible as DTL transformations, synthesize corresponding DTL programs, and use a cost model to determine when offloading the transformation is beneficial~\cite{HuangCompilingImpulse2001}.

There are also multiple avenues to integrate our framework into existing technologies such as MLIR~\cite{mlir}, where we envision DTU being used to provide materialization-free transformations into existing tensor framework cost models. Another is deeper integration into RISC-V, particularly targeting memory management methods to provide more flexibility to DTU users. We also believe that the flexibility of the DTU interface makes it a strong candidate for tight integration into a database query optimizer, where it could enable more effective optimization decisions.

Finally, while this work focuses on memory-centric data transformation, integrating DTU with compute-centric accelerators such as Gemmini~\cite{gemmini} is an interesting direction for future work. We believe DTU can complement such architectures by reducing data movement overheads and enabling more efficient data layouts prior to computation.

\section{Conclusion}\label{sec:conclusion}
This paper introduced DTL and DTU: a hardware–software co-design framework for generalized, on-the-fly data transformation. By enabling programmers to describe complex layout reorganizations in DTL, the DTU transparently bridges the gap between logical data structures and physical memory layouts. Our end-to-end evaluation on a RISC-V SoC shows that the DTU achieves substantial gains in execution speed and memory efficiency without requiring extensive modifications to computation kernels. Ultimately, this approach effectively decouples data layout from computation, enabling optimal access patterns while preserving software simplicity.


\section*{Acknowledgments}
This research is supported in part by NSF CCF-2403013 and NSF CCF-2403012.


\clearpage
\appendix


\section{DTL Grammar Specification}
This section presents the grammar specification of the Data Transformation Language (DTL), defining the syntax of valid kernels.

\begin{minipage}{\linewidth}
\begin{lstlisting}[caption={DTL Grammar Specification}, label={lst:appendix-grammar}]
<program> ::= <constdecls> <forstatement>

<constdecls> ::= <constdecls> <constdecl> | epsilon

<constdecl> ::= <opt> <type> ID ASSIGN INTLITERAL SEMICOL
              | <opt> <type> ID ASSIGN LCURLY <intlist> RCURLY SEMICOL

<opt> ::= NOPT | epsilon

<intlist> ::= INTLITERAL COMMA <intlist>
            | INTLITERAL
            | epsilon

<forstatement> ::= FOR LPAREN <constdecl> <id> LT <expr> SEMICOL 
                    <unarystmt> RPAREN LCURLY <forstatement> 
                    RCURLY
                    | FOR LPAREN <constdecl> <id> LT <expr>  
                    SEMICOL <unarystmt> RPAREN LCURLY <innernest> 
                    RCURLY

<innernest> ::= <outstatements>
                | <ifstatement>
                | <switchstatement>

<outstatements> ::= <outstatements> <outstatement>
                    | <outstatement>

<ifstatement> ::= IF LPAREN <ifpredicate> RPAREN LCURLY 
                  <outstatements> RCURLY ELSE LCURLY
                  <outstatements> RCURLY
                
<ifpredicate> ::= <id> <compareop> <id>
                  | ISEVEN id
                  | PAD <id> <id> <id> <id>
                  | ISEDGE <id>
                  | ISEDGE <id> OR <id>
                  | ISEDGE <id> AND <id>

<compareop> ::= LT | LTE | GT | GTE
<boolop> ::= AND | OR

<switchstatement> ::= SWITCH LPAREN <id> RPAREN LCURLY <casestatements> RCURLY

<casestatements> ::= <casestatement> <casestatements>
                     | <casestatement>

<casestatement> ::= CASE COLON <outstatements>

<outstatement> ::= OUT ASSIGN <expr> SEMICOL

<type> ::= INT

<expr> ::= <expr> <exprop> <expr>
           | < term>

<exprop> ::= CROSS | STAR | MINUS | LT

<term> ::= <factor>
           | LPAREN <expr> RPAREN

<factor> ::= INTLITERAL
             | <loc>

<loc> ::= <id>
          | <id> LBRACKET <id> RBRACKET

<id> ::= ID


\end{lstlisting}
\end{minipage}

\section{More DTL Kernel Examples}
In this section, we present further examples of DTL kernels to complement the earlier entries. 

\begin{minipage}{0.95\linewidth}
\begin{lstlisting}[caption={Example DTL  Kernel Performing Tensor Slicing over the inner two dimensions}, label={lst:slicing}]
int D4 = 64;
int D3 = 64;
int D2 = 64;
int D1 = 64;
int stride_d2 = D1;
int stride_d3 = D2*stride_d2;
int stride_d4 = D3*stride_d3;
int sel_factor_d1 = 8;
int sel_factor_d2 = 2;
int sel_factor_d3 = 1;
int sel_factor_d4 = 1;

// bounds equal to D4/sel_factor_d4
for (int i = 0; i < D4; i++)
{
    // bounds equal to D3/sel_factor_d3
    for (int j = 0; j < D3; j++)
    {
        // bounds equal to D2/sel_factor_d2
        for (int k = 0; k < 32; k++)
        {
            // bounds equal to D1/sel_factor_d1
            for (int l = 0; l < 8; l++)
            {
                out = l*(sel_factor_d1) + j*(stride_d2*sel_factor_d2) + k*(stride_d3*sel_factor_d3) + i*(stride_d4*sel_factor_d4);
            }
        }
    }
}
\end{lstlisting}

\begin{lstlisting}[caption={Example DTL Kernel Performing a 3 Column Projection}, label={lst:colproj-kernel}]
int col_offsets = {4, 32, 48};
int row_size = 64;
int ROW_COUNT = 500000;

for (int i = 0; i < ROW_COUNT; i++)
{
    for (int j = 0; j < 3; j++)
    {
        out = i*row_size + col_offsets[j];
    }
}
\end{lstlisting}

\begin{lstlisting}[caption={Example DTL Kernel Performing Vol2Col}, label={lst:vol2col-kernel}]
int depth = 16;
int height = 128;
int width = 128;
int channels = 3;
int KSIZE = 3;

for (int c = 0; c < channels; c++) {
    for (int kt = 0; kt < KSIZE; kt++) {
        for (int kh = 0; kh < KSIZE; kh++) {
            for (int kw = 0; kw < KSIZE; kw++) {
                for (int t = 0; t < depth-KSIZE+1; t++) {
                    for (int h = 0; h < height-KSIZE+1; h++) {
                        for (int w = 0; w < width-KSIZE+1; w++) 
                        {
                            out = (c*(depth*height*width) + w + kw)
                                    + ((t + kt)*(height*width)
                                    + (h + kh)*width);
                        }
                    }
                }
            }
        }
    }
}         
\end{lstlisting}
\end{minipage}\\

\newpage

\section{Artifact Appendix}
\subsection{Abstract}
This artifact provides the source code associated with the DTU system presented in this paper. It includes the modified FireSim/Chipyard system integration, DTU and AGU RTL implementations, the Data Transformation Language (DTL) implementation, and the benchmark suite used in the paper.

{\small
\begin{itemize}
  \item {\bf Program: } DTL benchmark suite and FireSim-based evaluation environment.
  \item {\bf Compilation: } RISC-V software toolchain, FireSim/Chipyard hardware build flow.
  \item {\bf Run-time environment: } FireSim/Chipyard-based RISC-V SoC simulation environment.
    \item {\bf Software: } Ubuntu 20.04-based host system, Xilinx Vivado
    \item {\bf Hardware: } Xilinx VCU118 FPGA
    \item {\bf Experiments: } Set up the Firesim environment, build the bitstream with the DTU-enabled config, compile and set up the software benchmark suite, run the benchmark suite in Firesim on the VCU118, and collect output logs in CSV format over the Firesim serial port. 
\item {\bf How much disk space required (approximately)?: } At least 300 GB
\item {\bf How much time is needed to prepare workflow (approximately)?: } 4-9 hours total. Firesim setup (2-3 hours). Bitstream compilation (2-6 hours). 
\item {\bf How much time is needed to complete experiments (approximately)?: } 6 hours
  \item {\bf Publicly Available?:} Yes.
\end{itemize}
}

\subsection{Description}

\subsubsection{How to access} All source code associated with the artifact is publicly available on GitHub. The artifact consists of a modified Firesim repository containing the system-level integration of the DTU, separate repositories containing the DTU and AGU RTL implementations, the DTL language implementation, and the DTL benchmark suite. Repository links are provided at: \begin{itemize}
    \item \href{https://github.com/CSL-KU/ProgrammableDataLayoutsASPLOS27}
               {\texttt{FireSim Fork}}

    \item \href{https://github.com/ColeStrickler/DTU/tree/5bbb64ca36b0850973d18ba049ca98a80d73d83d}
               {\texttt{Data Transformation Unit}}

    \item \href{https://github.com/ColeStrickler/DTL/tree/zero}
               {\texttt{Data Transformation Language}}

    \item \href{https://github.com/ColeStrickler/dtl_benchmarks/tree/DTU-CR}
               {\texttt{Benchmark Suite}}
\end{itemize}
For archival access and reproducibility, we also provide the following Zenodo DOIs: \begin{itemize}
    \item \href{https://doi.org/10.5281/zenodo.22308203}
               {\texttt{DTL Language/Compiler}}

    \item \href{https://doi.org/10.5281/zenodo.22308173}
               {\texttt{Data Transformation Unit}}

    \item \href{https://doi.org/10.5281/zenodo.22308190}
               {\texttt{Address Generation Unit}}

    \item \href{https://doi.org/10.5281/zenodo.22308194}
               {\texttt{Benchmark Suite}}
           
\end{itemize}\subsubsection{Hardware dependencies} Full-system evaluation requires a Xilinx VCU118 FPGA and a Linux-based host system. At least 300~GB of available disk space is recommended for the Firesim/Chipyard environment, FPGA build products, and benchmark artifacts. \subsubsection{Software dependencies} Firesim v1.18.0 is officially supported on Ubuntu 20.04. The evaluation environment additionally requires the Firesim/Chipyard software stack, a RISC-V cross-compilation toolchain, and Xilinx Vivado for FPGA bitstream generation. 

\subsection{Installation}

The artifact repositories contain installation and setup instructions for each component. Full-system evaluation requires installation of Firesim v1.18.0 and Chipyard on an Ubuntu 20.04 host, installation of the DTL toolchain and benchmark suite, and configuration of Xilinx Vivado for FPGA bitstream generation. Detailed installation commands, required software versions, and repository revisions are provided in the corresponding artifact repositories. A complete Firesim environment setup typically requires approximately 2--3 hours, excluding FPGA bitstream generation.

\end{document}